\documentclass[12pt,letterpaper]{article}

\usepackage{epsfig}

\begin{document}

\begin{center}
{\LARGE\bf Lattice fermions with complex mass}
\end{center}
\vspace{5pt}

\begin{center}
{\large\bf Stephan D\"urr$\,{}^{a}$}
\hspace{5pt}{\large and}\hspace{5pt}
{\large\bf Christian Hoelbling$\,{}^{b}$}
\\[10pt]
${}^a\,$ITP, Universit\"at Bern, Sidlerstr.\,5, CH-3012 Bern, Switzerland\\
${}^b\,$Bergische Universit\"at Wuppertal, Gaussstr.\,20, D-42119 Wuppertal,
Germany
\end{center}
\vspace{5pt}

\begin{abstract}
\noindent
We present evidence in the Schwinger model that rooted staggered fermions may
correctly describe the $m<0$ sector of a theory with an odd number of flavors.
We point out that in QCD-type theories with a complex-valued quark mass every
non-chiral action essentially ``borrows'' knowledge about the
$\theta$-transformation properties from the overlap action.
\end{abstract}

\newcommand{\pa}{\partial}
\newcommand{\pas}{\partial\!\!\!/}
\newcommand{\Dsl}{D\!\!\!\!/\,}
\newcommand{\hqu}{\hbar}
\newcommand{\ovr}{\over}
\newcommand{\til}{\tilde}
\newcommand{\pri}{^\prime}
\renewcommand{\dag}{^\dagger}
\newcommand{\<}{\langle}
\renewcommand{\>}{\rangle}
\newcommand{\gaf}{\gamma_5}
\newcommand{\lap}{\triangle}
\newcommand{\dal}{{\sqcap\!\!\!\!\sqcup}}
\newcommand{\trc}{\mathrm{tr}}
\newcommand{\Mpi}{M_\pi}
\newcommand{\Fpi}{F_\pi}

\newcommand{\al}{\alpha}
\newcommand{\be}{\beta}
\newcommand{\ga}{\gamma}
\newcommand{\de}{\delta}
\newcommand{\ep}{\epsilon}
\newcommand{\ve}{\varepsilon}
\newcommand{\ze}{\zeta}
\newcommand{\et}{\eta}
\renewcommand{\th}{\theta}
\newcommand{\vt}{\vartheta}
\newcommand{\io}{\iota}
\newcommand{\ka}{\kappa}
\newcommand{\la}{\lambda}
\newcommand{\rh}{\rho}
\newcommand{\vr}{\varrho}
\newcommand{\si}{\sigma}
\newcommand{\ta}{\tau}
\newcommand{\ph}{\phi}
\newcommand{\vp}{\varphi}
\newcommand{\ch}{\chi}
\newcommand{\ps}{\psi}
\newcommand{\om}{\omega}

\newcommand{\psb}{{\bar{\psi}}}
\newcommand{\etb}{{\bar{\eta}}}
\newcommand{\psh}{{\hat{\psi}}}
\newcommand{\eth}{{\hat{\eta}}}
\newcommand{\psd}{{\psi^{\dagger}}}
\newcommand{\etd}{{\eta^{\dagger}}}
\newcommand{\kh}{{\hat k}}
\newcommand{\qh}{{\hat q}}

\newcommand{\bdm}{\begin{displaymath}}
\newcommand{\edm}{\end{displaymath}}
\newcommand{\bea}{\begin{eqnarray}}
\newcommand{\eea}{\end{eqnarray}}
\newcommand{\beq}{\begin{equation}}
\newcommand{\eeq}{\end{equation}}

\newcommand{\mr}{\mathrm}
\newcommand{\mb}{\mathbf}
\newcommand{\Nf}{{N_{\!f}}}
\newcommand{\Nt}{{N_{t}}}
\newcommand{\Nc}{{N_{c}}}
\newcommand{\ri}{\mr{i}}
\newcommand{\Dov}{D^\mr{ov}}
\newcommand{\Dst}{D^\mr{st}}
\newcommand{\Dovm}{D^\mr{ov}_m}
\newcommand{\Dovam}{D^\mr{ov}_{|m|}}
\newcommand{\Dstm}{D^\mr{st}_m}
\newcommand{\Dstam}{D^\mr{st}_{|m|}}
\newcommand{\MeV}{\,\mr{MeV}}
\newcommand{\GeV}{\,\mr{GeV}}
\newcommand{\fm}{\,\mr{fm}}

\bigskip

The staggered fermion operator $\Dst$ describes $\Nt=2^{d/2}$ flavors
(``tastes'') of quarks in $d$ dimensions.
In order to obtain a single quark flavor in dynamical fermion simulations, it
has become customary to include the $\Nt$-th root of the staggered determinant
as a weight factor.

In a recent preprint \cite{Creutz:2006ys} Creutz has argued that this procedure
makes it difficult to obtain the right continuum limit.
His argument is based on the observation that the staggered determinant is
strictly a function of $m^2$, whereas the determinant of a true single flavor
theory also contains odd powers of $m$, arising from the unpaired chiral modes
of the Dirac operator.

In a reply \cite{Bernard:2006vv} Bernard et al.\ have noted that the rooted
determinant always corresponds to a theory with positive fermion mass,
regardless of the sign of the staggered mass term.
In addition, they point out that rooting is a non-analytic procedure.
Accordingly, they discuss how taking the continuum limit first can lead to odd
powers of $|m|$.

In this note we want to address how two observables, the chiral condensate and
the topological susceptibility, behave for non-positive quark masses.
We demonstrate in the Schwinger model \cite{Schwinger:1962tp}
that even with rooted staggered fermions these quantities can be measured for
$m<0$, by simply reinserting the phase information which gets lost by starting
with a doubled action and taking the square-root.
This ``recipe'' can be justified by reference to the overlap action, but
it is beyond the realm of a purely Lagrangian approach with staggered fermions.

The numerical data presented below are based on a sample of $1000$ gauge
configurations at $\be=9.8$ on a $N\!\times\!N=28^2$ lattice.
We compare (rooted) staggered to overlap fermions \cite{Neuberger:1997fp},
both actions defined with one step of APE smearing.
For details see \cite{Durr:2003xs,Durr:2004ta}.
Throughout, we use the coupling $e$ to define the physical scale and set the
lattice spacing to $a=1$.

\bigskip 

First, we consider the scalar condensate.
We are interested in its continuum limit in a fixed physical volume (cf.\
\cite{Durr:2004ta}), and for $\Nf\!=\!1$ its presence is due to the anomaly
\cite{Schwinger:1962tp} and does thus not signal spontaneous symmetry breaking.
For overlap fermions and $m>0$ its bare version is
\beq
{\ch_\mr{scal}^\mr{ov}\ovr e}={\sqrt{\be}\ovr N^2}
\frac{\<\det(\Dovm)^\Nf\sum'{1\ovr\til\la+m}\>}{\<\det(\Dovm)^\Nf\>}
\;,\qquad
\til\la={1\ovr{1\ovr\la}-{1\ovr2\rh}}
\label{condover}
\eeq
where the sum runs over all eigenvalues of the massless operator $\Dov$, except
for the doublers (on topologically non-trivial backgrounds) at $2\rho$.
Here, $\rh$ is a parameter in the overlap construction which in our runs
is set to $1$.
The measure in (\ref{condover}) is quenched with an explicit weight factor
\beq
\det(\Dovm)=\prod((1-{m\ovr 2\rh})\la+m)
\;.
\label{detover}
\eeq
For staggered fermions and $m>0$ the definition of the bare condensate is
analogous
\beq
{\ch_\mr{scal}^\mr{st}\ovr e}={\sqrt{\be}\ovr 2N^2}
\frac{\<\det(\Dstm)^{\Nf/2}\sum{1\ovr\la+m}\>}{\<\det(\Dstm)^{\Nf/2}\>}
\label{condstag}
\eeq
except that the sum over all eigenvalues plus the prefactor $2^{-1}=\Nt^{-1}$
provide a taste projection in the valence sector and the ensemble average is
taken with the square-root of the determinant
\beq
\det(\Dstm)=\prod(\la+m)
\label{detstag}
\eeq
to correct for the extra degree of freedom in the sea sector.
For $m>0$ the condensates (\ref{condover}, \ref{condstag}) have a
logarithmically divergent contribution, and after subtracting the latter, they
seem to yield a universal quantity in the continuum, regardless whether $\Nf=1$
or $\Nf=2$ (where there is no rooting issue in 2D) \cite{Durr:2004ta}.
Below, this subtraction is always implicit.

\begin{figure}
\centerline{\epsfig{file=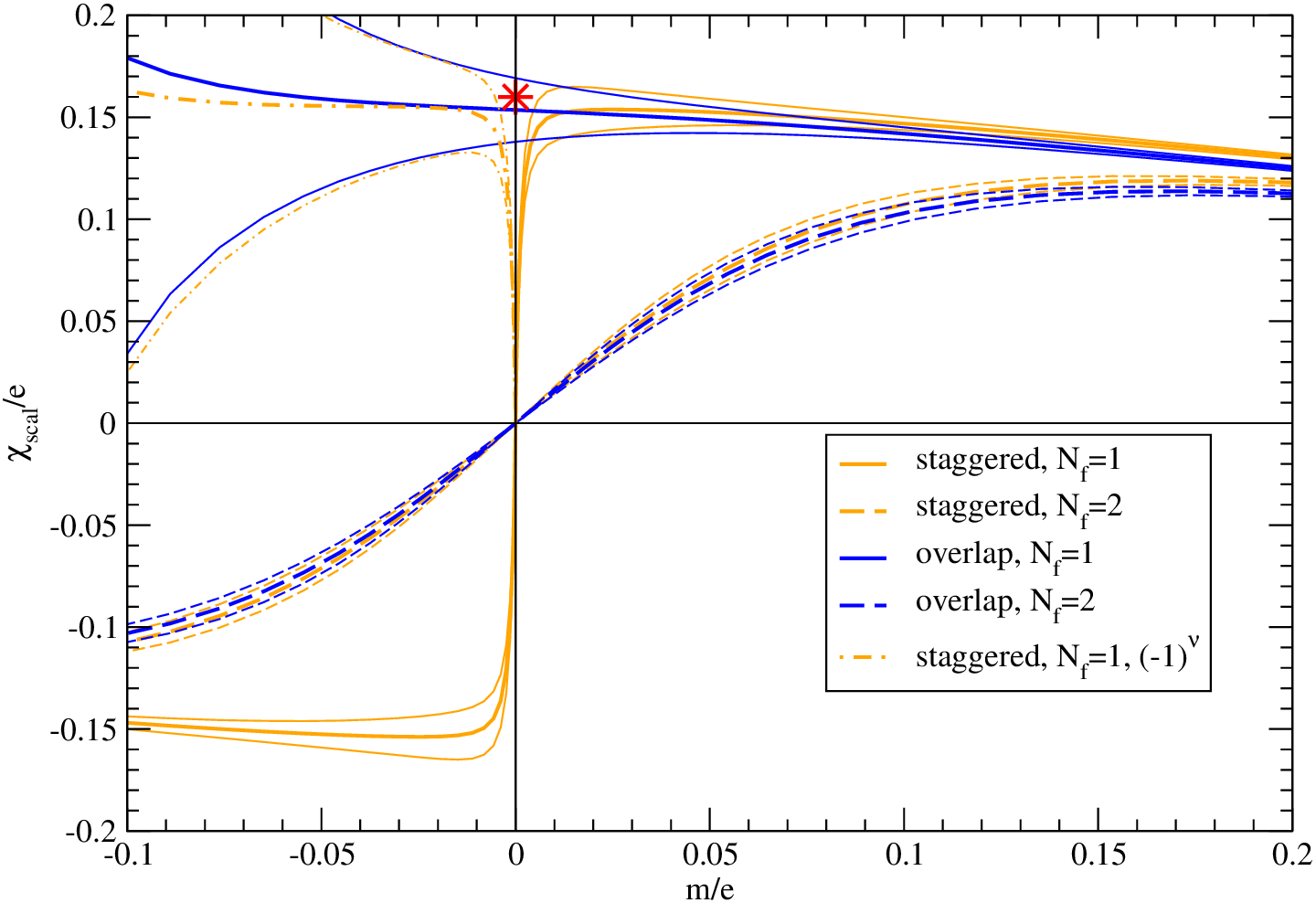,width=14cm}}
\vspace{-2mm}
\caption{Scalar condensate in the $\Nf=1,2$ dynamical theory versus quark mass.
For $\Nf=1$ and $m=0$ the overlap condensate successfully reproduces the
analytic value by Schwinger. For $\Nf=1$ and $m<0$ only the ``smart'' staggered
curve [with an explicit factor $(-1)^\nu$] is correct.}
\label{fig_cond}
\vspace{6mm}
\centerline{\epsfig{file=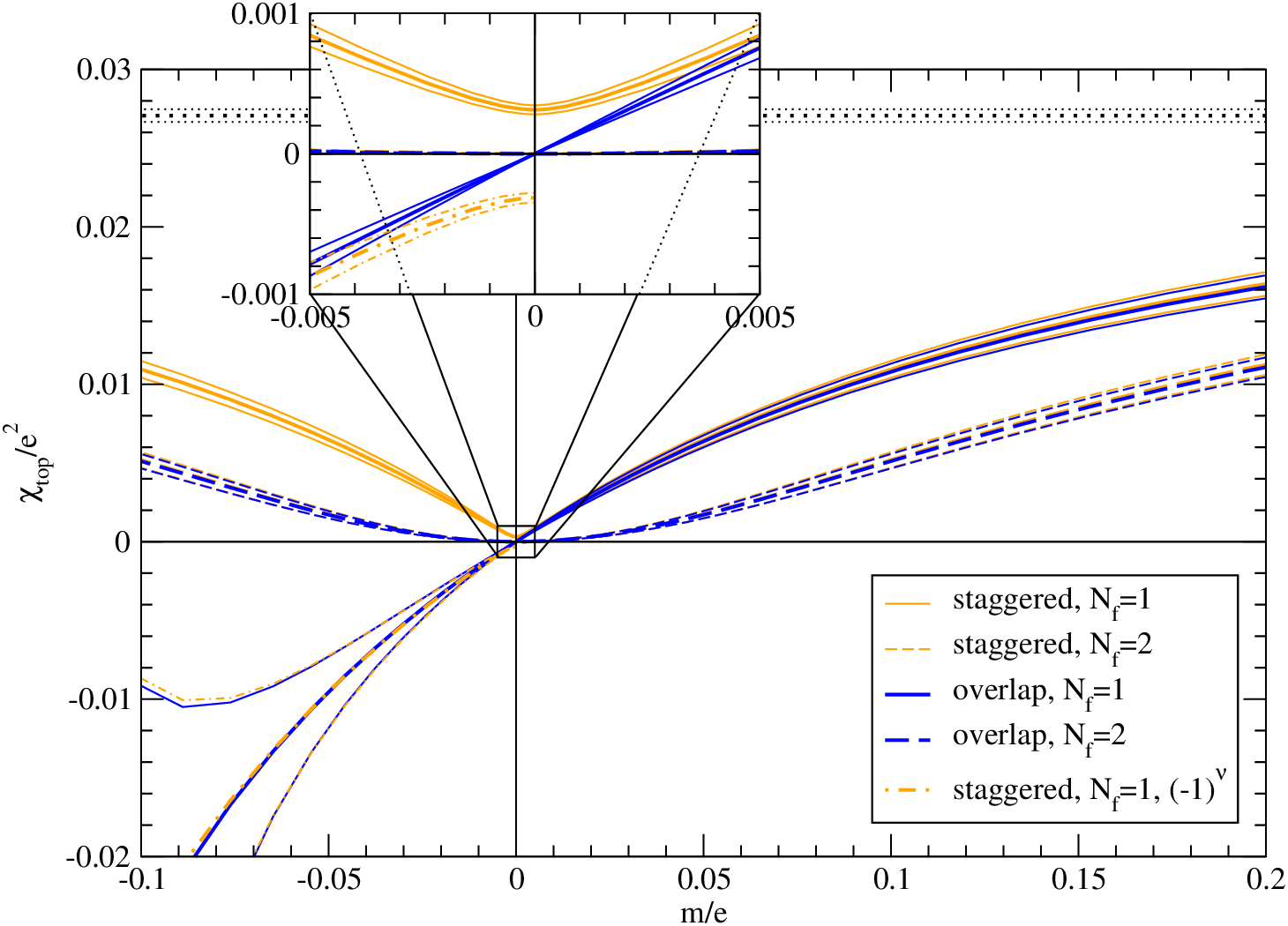,width=14cm}}
\vspace{-2mm}
\caption{Topological susceptibility in the $\Nf=1,2$ dynamical theory versus
quark mass. For $\Nf=1$ and $m<0$ only the ``smart'' staggered curve [with an
explicit factor $(-1)^\nu$] is correct. The quenched topological susceptibility
is included to indicate the asymptotic $m\to+\infty$ value.}
\label{fig_susc}
\end{figure}

We now evaluate (\ref{condover}, \ref{condstag}) for arbitrary real $m$
with details as specified in \cite{Durr:2004ta}.
Triple curves in Fig.\,\ref{fig_cond} indicate $1\si$ error bands.
Errors are correlated, since the same gauge configurations have been used.
We first discuss the $\Nf=2$ case.
Here, our overlap and staggered data are in good agreement.
They are both odd in $m$, as expected with an even partition function and one
derivative w.r.t.\ the quark mass.
The $\Nf=1$ case is more subtle.
Here, our overlap data are smooth near $m=0$, reproducing the analytic result
$\ch_\mr{scal}/e=\exp(\ga)/(2\pi^{3/2})=0.1599...$ by Schwinger
\cite{Schwinger:1962tp}.
On the other hand, the rooted staggered data pass through the origin, as they
have to, since their partition function is even in $m$, and a derivative
w.r.t.\ the quark mass yields an odd function.
The 1-flavor overlap curve is neither even nor odd, since the operator being
odd is not sufficient for inducing this property in the observable.
The functional measure is even in $m$ with a straightforward (``naive'')
evaluation of (\ref{condstag}, \ref{detstag}) in case of staggered fermions.
It is, however, not even with an odd number of overlap fermions, as is obvious
from the fact that on a background with index $\nu$ the overlap determinant has
the same sign as $m^{\Nf|\nu|}$.

With this information at hand, it is now easy to understand how the failure of
the ``naive'' staggered rooting procedure at $m<0$ could be corrected.
What is missing is the same information that gets lost if one first squares the
overlap determinant and then takes the square-root.
This suggests that -- rather than naively evaluating the product in
(\ref{detstag}) for any $m$ -- one should define the functional determinant
of a rooted staggered fermion at negative mass via
\beq
\det(\Dst,\Nf=1,-|m|)\equiv(-1)^{|\nu|}\det(D^\mr{st}_{|m|})^{1/\Nt}
\label{detnegstag}
\eeq
in close analogy to $\,\det(\Dov_{-|m|})=(-1)^{|\nu|}\det(\Dov_{|m|})\,$
(see below).
We have tried the definition (\ref{detnegstag}), together with a sign-flip in
the condensate (see below) for $\Nf=1$; the result is shown with dash-dotted
lines in Fig.\,\ref{fig_cond}.
This definition lets the ``smart'' staggered action reproduce the overlap
answer for the 1-flavor condensate at (sufficiently) negative mass.
Together with the standard definition at positive mass, the staggered answer is
thus qualitatively correct, except for a small dip at $m=0$.
This dip gets narrowed (horizontally), as the coupling tends to zero
($\be\to\infty$), with details given in \cite{Durr:2004ta}.
Still, since the value at $m=0$ remains an exact zero at any $\be$, the
continuum extrapolated staggered curve is discontinuous at this point.
This shows that in general the limits $m\to0$ and $a\to0$ do not commute with
staggered fermions \cite{Durr:2004ta,Bernard:2004ab}.

\bigskip 

Next, we consider the topological susceptibility in the same finite box.
It is defined through
\beq
{\ch_\mr{top}\ovr e^2}={\be\ovr N^2}
{\<\det(D_m)^\Nf\;\nu^2\>
\ovr
\<\det(D_m)^\Nf\>}
\label{topsusc}
\eeq
and we choose to evaluate the index $\nu$ with the overlap operator, but other
options would just bring different cut-off effects.
The determinant in (\ref{topsusc}) is meant to represent the eigenvalue product
on the r.h.s.\ of (\ref{detover}) for overlap fermions.
For staggered quarks there is the ``naive'' option to use the r.h.s.\ of
(\ref{detstag}) for all $m$, versus the ``smart'' one to use it only for $m>0$,
and to continue to $m<0$ via (\ref{detnegstag}).

Our results for the topological susceptibility with real $m$ are shown in
Fig.\,\ref{fig_susc}.
For $\Nf=2$ both actions are in perfect agreement; in particular both curves
are strictly even in $m$.
For $\Nf=1$ the overlap curve goes right through the origin, but it is not
exactly odd in $m$.
It is negative for $m<0$, in spite of the observable $\nu^2$ being positive
or zero, simply because the determinant may have either sign.
By contrast, the ``naive'' staggered curve is still exactly even and positive.
Being smooth, it must show a residual susceptibility at $m=0$, a pure
cut-off effect which tends to zero, under $\be\to\infty$, with a
rather high power of $ae$ \cite{Durr:2004ta}.
Again, defining a ``smart'' staggered action at $m<0$ via (\ref{detnegstag})
basically flips the sign of the staggered 1-flavor curve (for small enough
$|m|$) and makes it agree with the overlap data.
Still, the derivative from the right, evaluated at $m=0$, is a ``malicious''
observable where the staggered continuum limit is not correct.
It is always zero, whereas the overlap counterpart indicates that the
true continuum-limit is non-vanishing.
This is another example of the non-commutativity of the $m\to0$ and $a\to0$
limits in the $\Nf=1$ theory \cite{Durr:2004ta}.
Regarding the $m>0$ piece of the curve, we find it most amusing that the
staggered version consists of a rather large tower of even (in $m$)
contributions which --~all together~-- manage to mimic an almost \emph{linear}
behavior.


We add a remark regarding $m<0$.
Our graphs in Figs.\,\ref{fig_cond},\,\ref{fig_susc} are cut off at $m/e=-0.1$.
This is because the sign problem in the 1-flavor theory hits us hard to the
left of this point, and the error-bars explode.
Note that there is no sign-problem in the 2-flavor theory, regardless of the
action, and the ``naive'' staggered 1-flavor determinant does thus not induce
a sign problem.
By contrast, the ``smart'' staggered 1-flavor theory has the same sign
problem as the overlap version, and this speaks in favor of (\ref{detnegstag})
being an honest approach to the $m<0$ region.

\bigskip 

We now extend the discussion to QCD with arbitrary complex quark mass,
assuming that the fundamental $\th$-term in the action is zero (if not, it
would entail a trivial generalization).

In the continuum, a complex quark mass $m=|m|e^{\ri\th}$ can be rotated into
such a $\th$-term, and the derivation builds on chiral symmetry
\cite{Leutwyler:1992yt}.
In a lattice regulated theory the same holds true if the lattice action
respects exact chiral symmetry.
To this end it is useful to note that the standard definition of the massive
overlap operator, $\Dovm=(1-{m\ovr2\rh})\Dov+m$, is not adequate for complex
$m$.
For reasons to become obvious, we represent the action in the general case as
\beq
S^\mr{ov}_m=
\psb_+\Dov\psh_+ + \psb_-\Dov\psh_- + m\psb_-\psh_+ + m^*\psb_+\psh_-
\label{over_ori}
\eeq
(compare \cite{Leutwyler:1992yt} and \cite{Niedermayer:1998bi})
where the positive/negative chirality fields are defined through
\beq
\psh_\pm=\hat{P}_\pm\ps,\quad
\psb_\pm=\psb P_\mp,\qquad
\eeq
\beq
\hat{P}_\pm={1\ovr2}(1\pm\hat\gaf),\quad
P_\pm={1\ovr2}(1\pm\gaf),\qquad
\hat\gaf=\gaf(2\til1-1),\quad\til1=1-{1\ovr2\rh}\Dov
\eeq
and hat/tilde-quantities implicitly depend on the gauge field.
By defining rotated fields
\beq
\psh_\pm'=e^{\pm\ri\th/2}\psh_\pm,\quad
\psb_\pm'=\psb_\pm e^{\mp\ri\th/2}
\label{trafo}
\eeq
the overlap action (\ref{over_ori}) can be brought into the form
\beq
(S^\mr{ov}_m)'=
\psb_+'\Dov\psh_+' + \psb_-'\Dov\psh_-' + |m|\psb_-'\psh_+' + |m|\psb_+'\psh_-'
\label{over_tra}
\eeq
which reflects only the modulus of the quark mass.
In addition, one has to take into account the anomalous transformation of the
1-flavor path-integral measure \cite{Fujikawa:1980eg}
\beq
D\psb D\ps\longrightarrow
D\psb' D\ps'=D\psb D\ps\,\det\!\Big({\de\psb'\,\de\ps'\ovr\de\psb\,\de\ps}\Big)
\eeq
and with $\de\psb'/\de\psb=e^{\ri\gaf\th/2}$,
$\de\ps'/\de\ps=e^{\ri\hat\gaf\th/2}$ and the
index theorem \cite{Hasenfratz:1998ri} it follows that
\beq
\det\!\Big({\de\psb'\,\de\ps'\ovr\de\psb\,\de\ps}\Big)=
e^{\ri\,\mr{tr}(\gaf+\hat\gaf)\th/2}=
e^{-\ri\,\mr{tr}(\gaf\Dov)\th/(2\rh)}=
e^{\ri\nu\th}\;.
\eeq
With a chirally symmetric action one can thus rotate the effect of a complex
mass into a term
\beq
S_\th=-\ri\nu\th
\label{theta}
\eeq
in the action and stay with a positive fermion mass $|m|$.
In a theory with $\Nf$ degenerate overlap fermions it is convenient to
parameterize each mass as $m=|m|e^{\ri\th/\Nf}$, since this choice again yields
the $\th$-term (\ref{theta}).
It is straightforward to generalize the discussion to an arbitrary complex
$\Nf\times\Nf$ mass matrix, and the result is the same if one lets $\th$
represent the phase of the determinant of this matrix
\cite{Leutwyler:1992yt}.
The important consequence of the $\th$-term (\ref{theta}) reflecting the
anomalous transformation behavior of the QCD path-integral measure
is that the full theory is periodic under a simultaneous $2\pi/\Nf$ shift in
the phase of each quark mass.

With staggered fermions it makes no sense to ``naively'' evaluate the action
with a complex mass.
However, the discussion in the overlap case has shown that it is sufficient to
have an explicit $\th$-term (\ref{theta}) and a fermion action with a positive
mass.
Hence, if we \emph{assume} that the $\Nf/\Nt$-th root of the staggered
determinant yields the right continuum limit for $\Nf$ positive-mass fermions,
then it is perfectly reasonable to stay with that root and to introduce an
explicit $\th$-term (\ref{theta}) in the action.
The pertinent $2\pi/\Nf$ period under a common shift in the phase of each quark
mass in the resulting continuum theory is no further assumption.

In order to compute observables for arbitrary complex mass via the $\th$-term
prescription, one has to perform a chiral rotation in the valence sector, too.
We first discuss the overlap case.
Representing the mass term in (\ref{over_ori}) as
$|m|\psb(P_+e^{\ri\th}\hat{P}_++P_-e^{-\ri\th}\hat{P}_-)\ps$
and using the identities
\bea
P_+\hat{P}_+&=&
{1\ovr2}(1+\gaf)(1-{1\ovr2\rh}\Dov)={1\ovr2}(1+\gaf)\til1
\\
P_-\hat{P}_-&=&
{1\ovr2}(1-\gaf)(1-{1\ovr2\rh}\Dov)={1\ovr2}(1-\gaf)\til1
\eea
the 1-flavor action with a complex mass $m=|m|e^{\ri\th}$ can be written in the
compact form
\beq
S^\mr{ov}_m=\psb\Dovm\ps\;,\quad
\Dovm\equiv\Dov+|m|e^{\ri\gaf\th}\til1
\;.
\label{over_new}
\eeq
In the theory with complex mass $m=|m|e^{\ri\th}$ and vanishing $\th$-term,
the (unsubtracted and unrenormalized) scalar/pseudoscalar condensate is given
by
\bea
\ch_\mr{scal}&=&
\;-\<\psb\til1\ps\>\;=\;
-\<\psb(P_+\hat{P}_++P_-\hat{P}_-)\ps\>=
-\<\psb_-\psh_++\psb_+\psh_-\>
\label{chis_ov}
\\
\ch_\mr{pseu}&=&
-\<\psb\gaf\til1\ps\>=
-\<\psb(P_+\hat{P}_+-P_-\hat{P}_-)\ps\>=
-\<\psb_-\psh_+-\psb_+\psh_-\>
\label{chip_ov}
\eea
if one includes the determinant of the massive operator (\ref{over_new}) into
the functional measure.
We now express them in terms of the primed fields which refer to the theory
with positive mass
\bea
\ch_\mr{scal}&=&
-\<\psb_-\psh_++\psb_+\psh_-\>=
-\<\psb_-'e^{-\ri\th}\psh_+'+\psb_+'e^{\ri\th}\psh_-'\>=
\cos(\th)\,\ch_\mr{scal}'-\ri\sin(\th)\,\ch_\mr{pseu}'
\nonumber
\\
\ch_\mr{pseu}&=&
-\<\psb_-\psh_+-\psb_+\psh_-\>=
-\<\psb_-'e^{-\ri\th}\psh_+'-\psb_+'e^{\ri\th}\psh_-'\>=
\cos(\th)\,\ch_\mr{pseu}'-\ri\sin(\th)\,\ch_\mr{scal}'
\nonumber
\eea
where the latter quantities adopt a simple form which shows invariance
under chiral rotations
\beq
\ch_\mr{scal}'=
\<\mr{Tr}[\til1(\Dov+|m|\til1)^{-1}]\>=
{1\ovr V}\<\sum\nolimits'{1\ovr\til\la\!+\!|m|}\>
\,,\;
\ch_\mr{pseu}'=
\<\mr{Tr}[\gaf\til1(\Dov+|m|\til1)^{-1}]\>=
{\<\nu\>\ovr V|m|}
\,.\!\!
\label{chibothprime}
\eeq
In these formulas the functional measure is still the one in the theory with
the complex mass.
Our previous exercise tells us that this is equivalent to the quenched measure
with $\det(\Dovam)$ and an additional factor $e^{\ri\nu\th}$.
In other words, we need to calculate the expressions in (\ref{chibothprime})
times some reweighting factors on quenched backgrounds.
In a theory with $\Nf$ degenerate flavors of mass $m=|m|e^{\ri\th/\Nf}$ the
scalar/pseudoscalar condensate (per flavor) is thus given by
\bea
\ch_\mr{scal}^\mr{ov}&=&
{1\ovr V}\,{\<\det(\Dovam)^\Nf\;e^{\ri\nu\th}\;
[\cos(\th/\Nf)\sum'{1\ovr\til\la+|m|}-\ri\sin(\th/\Nf){\nu\ovr|m|}]
\>\ovr
\<\det(\Dovam)^\Nf\;e^{\ri\nu\th}\>}
\label{scalov}
\\
\ch_\mr{pseu}^\mr{ov}&=&
{1\ovr V}\,{\<\det(\Dovam)^\Nf\;e^{\ri\nu\th}\;
[\cos(\th/\Nf){\nu\ovr|m|}-\ri\sin(\th/\Nf)\sum'{1\ovr\til\la+|m|]}
\>\ovr
\<\det(\Dovam)^\Nf\;e^{\ri\nu\th}\>}
\label{pseuov}
\eea
where the sum is over all modes of the massless theory, except for the chiral
doublers, and $\til\la$ relates to $\la$ as indicated in (\ref{condover}).
With staggered fermions there is no alternative to defining the
scalar/pseudoscalar condensate (per taste) through the expression
\bea
\ch_\mr{scal}^\mr{st}&=&
{1\ovr V}\,{\<\det(\Dstam)^{\Nf/\Nt}\;e^{\ri\nu\th}\;
[\cos(\th/\Nf)\Nt^{-1}\sum{1\ovr\la+|m|}-\ri\sin(\th/\Nf){\nu\ovr|m|}]
\>\ovr
\<\det(\Dstam)^{\Nf/\Nt}\;e^{\ri\nu\th}\>}
\label{scalst}
\\
\ch_\mr{pseu}^\mr{st}&=&
{1\ovr V}\,{\<\det(\Dstam)^{\Nf/\Nt}\;e^{\ri\nu\th}\;
[\cos(\th/\Nf){\nu\ovr|m|}-\ri\sin(\th/\Nf)\Nt^{-1}\sum{1\ovr\la+|m|}]
\>\ovr
\<\det(\Dstam)^{\Nf/\Nt}\;e^{\ri\nu\th}\>}
\label{pseust}
\eea
with the sum running over all modes of the massless theory and the factor
$\Nt^{-1}$ taking care of the degeneracy.
The case of the topological susceptibility is simpler.
Since the topological charge is invariant under a chiral rotation, we have in
the overlap case the identity
\beq
\ch_\mr{top}^\mr{ov}={1\ovr V}\,
{\<\det(\Dovam)^\Nf\;e^{\ri\nu\th}\;\nu^2\>\ovr
\<\det(\Dovam)^\Nf\;e^{\ri\nu\th}\>}
\label{topsuscthetaov}
\eeq
and the analogous definition for staggered fermions reads
\beq 
\ch_\mr{top}^\mr{st}={1\ovr V}\,
{\<\det(\Dstam)^{\Nf/\Nt}\;e^{\ri\nu\th}\;\nu^2\> \ovr
\<\det(\Dstam)^{\Nf/\Nt}\;e^{\ri\nu\th}\>}
\;.
\label{topsuscthetast}
\eeq
In all these expressions one may replace
$e^{\ri\nu\th}\to\cos(\nu\th),\ri\sin(\nu\th)$, depending on whether the
remainder is even or odd in $\nu$ (due to CP symmetry in the Yang-Mills theory).
We stress that an attempt to directly implement (\ref{over_ori}) with
Wilson fermions is conceptually not very much different from the approach
where one ``borrows'' expressions (\ref{pseuov}, \ref{topsuscthetaov})
from an action with exact flavor and chiral symmetry, as we have advocated
for staggered fermions.

\begin{figure}
\centerline{\epsfig{file=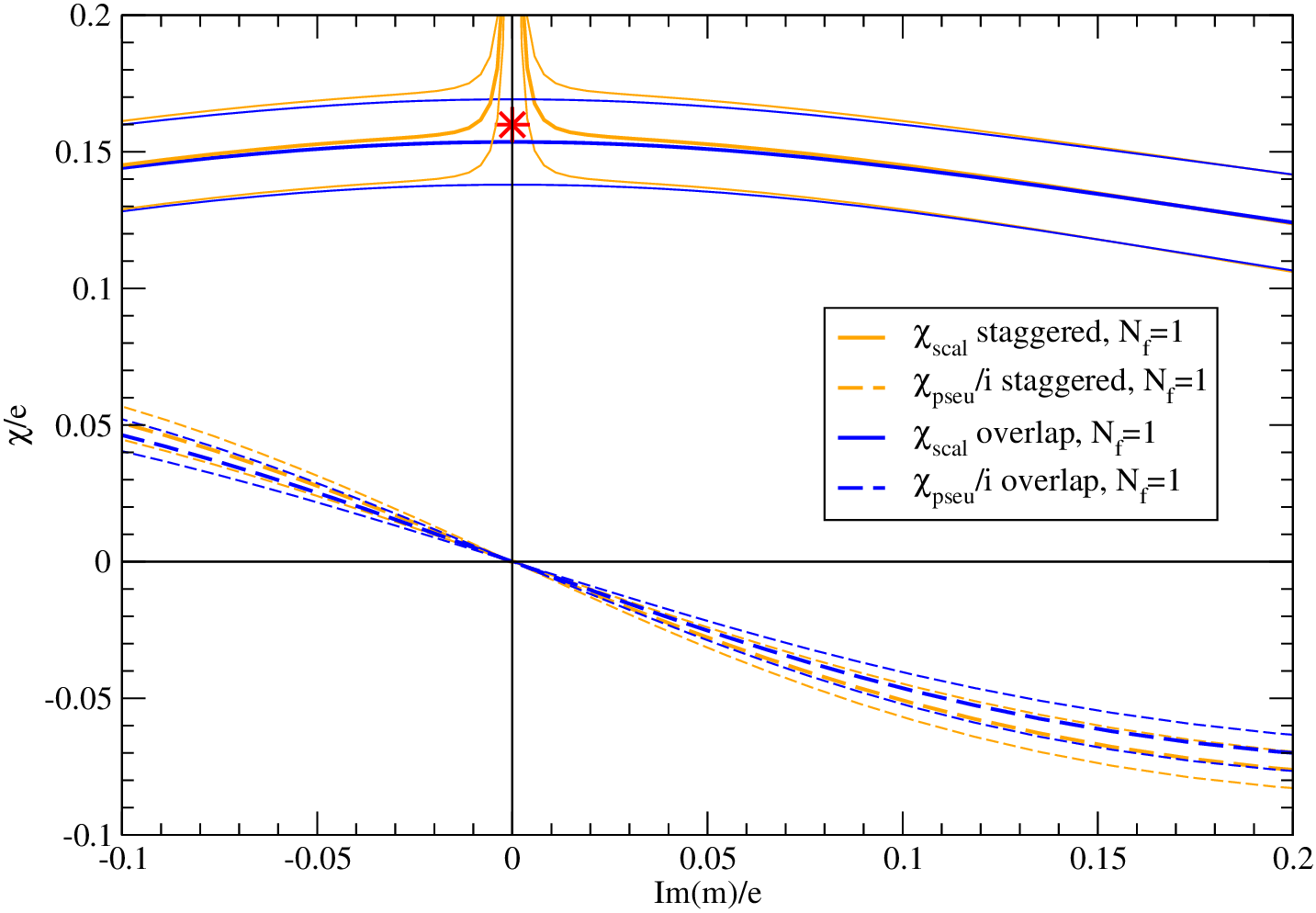,width=14cm}}
\caption{Scalar condensate and imaginary part of the pseudoscalar condensate
versus imaginary mass in the $\Nf=1$ dynamical theory, with the definition
(\ref{scalst}, \ref{pseust}) in the staggered case.}
\label{fig_condi}
\vspace{10mm}
\centerline{\epsfig{file=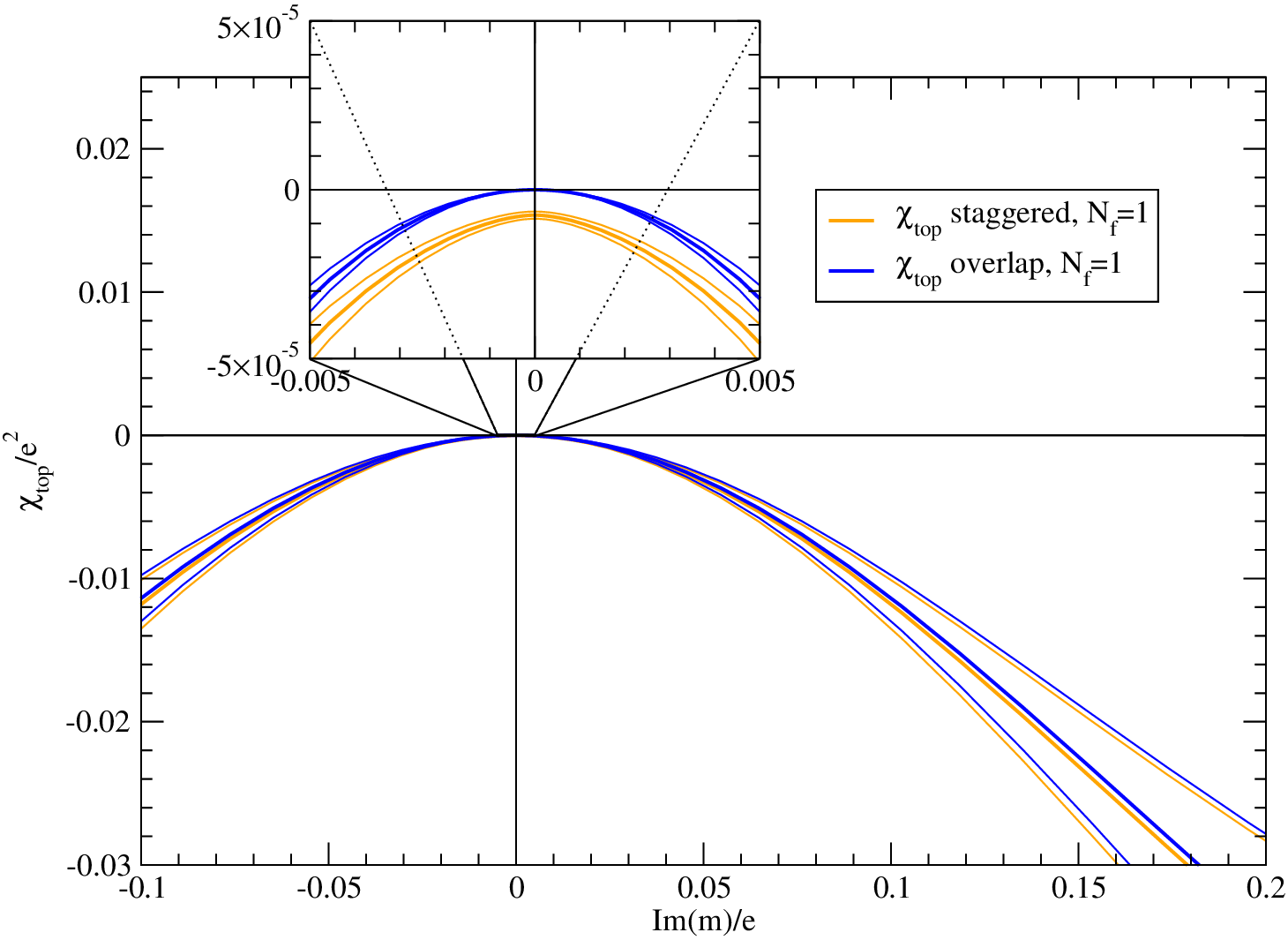,width=14cm}}
\caption{Topological susceptibility versus imaginary mass in the $\Nf=1$
dynamical theory. In the staggered case the definition (\ref{topsuscthetast})
has been used.}
\label{fig_susci}
\end{figure}

In order to test how this reduction works in practice we return to the $\Nf=1$
Schwinger model.
We restrict ourselves to $\th=\pm\pi/2$, i.e.\ to the imaginary mass axis.
Here, the scalar condensate is real, and the pseudoscalar one is purely
imaginary; $\ch_\mr{scal},\ch_\mr{pseu}/\ri$ are shown in Fig.\,\ref{fig_condi}.
For $\ch_\mr{scal}$, the overlap expression (\ref{pseuov}) shows a smooth
behavior, reproducing the Schwinger value in the $m\to0$ limit.
The staggered prescription (\ref{pseust}) seems consistent, except for a region
around $m=0$, where the non-commutativity phenomenon again occurs.
The absence of an exact matching of the zero-modes in the determinant and in
the observable leads to a blow-up in the staggered answer, if the chiral limit
is reached from the imaginary-mass direction.
In $\ch_\mr{pseu}/\ri$, the deviation between the two actions seems to be a
mild $O(a^2)$ cut-off effect.

Our data for the topological susceptibility in the $\Nf=1$ Schwinger model with
quark mass $m=|m|e^{\ri\th}$ and $\th=\pm\pi/2$ are shown in
Fig.\,\ref{fig_susci}.
In this case the situation looks entirely inconspicuous, and we expect that
there is a common continuum limit.

\bigskip 

We have demonstrated in a simple toy model that one may obtain sensible physics
at negative quark mass for an odd number of staggered continuum flavors, if one
uses a ``smart'' definition of the rooted staggered determinant which takes
into account the phase factor that got lost in the rooting procedure.
Without the advent of exact lattice chiral symmetry, this would be hard to
justify, since with staggered fermions it is definitely beyond what a purely
Lagrangian approach can do.
We close with three conjectures on staggered fermions:
\begin{enumerate}
\itemsep-1mm
\item
For $m>0$, taste-projected correlators built from $D_{\mr{st},m}^{-1}$
and $\det(D_{\mr{st},m})^{\Nf/\Nt}$ in concert yield the
\emph{correct continuum limit} for $d=2,4$, arbitrary gauge group and $\Nf$.
\item
For $m=0$, taste-projected correlators built from $D_{\mr{st},0}^{-1}$
and $\det(D_{\mr{st},0})^{\Nf/\Nt}$ define a theory in the
\emph{wrong universality class} (i.e.\ at least one observable assumes an
incorrect continuum value) for $d=2,4$, arbitrary gauge group and $\Nf$.
\item
The ``smart'' staggered determinant which effectively shifts the
$\th$-dependence of the theory with $m=|m|e^{\ri\th/\Nf}$ into an explicit
$\th$-term (\ref{theta}) yields the correct $\th$-dependence if and only if
the staggered continuum limit with real positive mass $|m|$ is correct.
\end{enumerate}

\bigskip 

{\bf Acknowledgments}:
We acknowledge E-mail exchanges with Claude Bernard and Steve Sharpe.
SD has benefited from the discussions during the ``workshop on the fourth root
of the staggered fermion determinant'' at the INT in Seattle, and from the
hospitality extended by this institution while completing this work.
CH acknowledges discussions with Daniel Nogradi.
This work was supported by the Swiss NSF, and partly by the United States DOE.

\end{document}